\documentclass[preprint,12pt]{elsarticle}

\usepackage{verbatim} 
\usepackage{epsfig}
\usepackage{amssymb} 
\usepackage{mathtools}
\usepackage{lineno}

\usepackage{tikz}        
\usepackage{etoolbox}

\usepackage[caption=false]{subfig}

\begin{document}

\newcommand*{\hwplotB}{\raisebox{3pt}{\tikz{\draw[red,dashed,line 
width=3.2pt](0,0) -- 
(5mm,0);}}}

\newrobustcmd*{\mydiamond}[1]{\tikz{\filldraw[black,fill=#1] (0,0) -- 
(0.1cm,0.2cm) --
(0.2cm,0) -- (0.1cm,-0.2cm);}}

\newrobustcmd*{\mytriangleleft}[1]{\tikz{\filldraw[black,fill=#1] (0,0.15cm) -- 
(-0.3cm,0) -- (0,-0.15cm);}}
\definecolor{Blue}{cmyk}{1.,1.,0,0} 

\begin{frontmatter}

\title{A re-examination of the role of friction in the original Social Force Model}

\author[label1]{I.M.~Sticco}
\author[label3]{G.A.~Frank}
\author[label1]{F.E.~Cornes}
\author[label1,label2]{C.O.~Dorso}
\address[label1]{Departamento de F\'\i sica, Facultad de Ciencias 
Exactas y Naturales, Universidad de Buenos Aires, Pabell\'on I, Ciudad 
Universitaria, 1428 Buenos Aires, Argentina.}
\address[label2]{Instituto de F\'\i sica de Buenos Aires, Pabell\'on I, Ciudad 
Universitaria, 1428 Buenos Aires, Argentina.}
\address[label3]{ Unidad de Investigaci\'on y Desarrollo de las 
Ingenier\'\i as, Universidad Tecnol\'ogica Nacional, Facultad Regional Buenos 
Aires, Av. Medrano 951, 1179 Buenos Aires, Argentina.}

\begin{abstract}
The fundamental diagram of pedestrian dynamics gives the relation between the density and the flow within a specific enclosure. It is characterized by two distinctive behaviors: the free-flow regime (for low densities) and the congested regime (for high densities). In the former, the flow is an increasing function of the density, while in the latter, the flow remains on hold or decreases. In this work, we perform numerical simulations of the pilgrimage at the entrance of the Jamaraat bridge (pedestrians walking along a straight corridor) and compare flow-density measurements with empirical measurements made by Helbing \textit{et al}~\cite{helbing3}. We show that under high density conditions, the basic Social Force Model (SFM) does not completely handle the fundamental diagram reported in empirical measurements. We use analytical techniques and numerical simulations to prove that with an appropriate modification of the friction coefficient (but sustaining the SFM) it is possible to attain behaviors which are in qualitative agreement with the empirical data. Other authors have already proposed a modification of the relaxation time in order to address this problem. In this work, we unveil the fact that our approach is analogous to theirs since both affect the same term of the reduced-in-units equation of motion. We show how the friction modification affects the pedestrian clustering structures throughout the transition from the free-flow regime to the congested regime. We also show that the speed profile, normalized by width and maximum velocity yields a universal behavior regardless of the corridor dimensions. \\
\end{abstract}


\end{frontmatter}


\section{\label{introduction}Introduction}

In the last decades, many microscopic models 
for crowd dynamics were developed. Force-based models were introduced by Hirai 
and Tarui \cite{Hirai} who in 1975 proposed to associate the interaction between 
pedestrians and their environment using different kinds of forces inspired by physical systems. For 
example, Okazaki (1979) modeled the movement of pedestrians as 
a magnetized object immersed a magnetic in field \cite{Okazaki}.\\

By the late 90's and the beginning of the century, Helbing 
\textit{et al.} proposed a model that included socio-psychological and physical 
forces to simulate crowd dynamics. According to Helbing \textit{et al.}, both 
the environment and the individuals' own desire affect the pedestrians motion in 
a similar way as forces do with respect to the momentum of particles 
\cite{Helbing1,Helbing4}. This ``Social Force Model'' (SFM) related the 
socio-psychological phenomenon of crowds behavior to the ``microscopic'' 
formalism of moving particles. The model succeeded in reproducing
the reduction of the efficiency of the evacuation process as 
pedestrians try harder to escape from a dangerous situation (\textit{i.e.} 
``faster is slower'' effect) \cite{Helbing1,Dorso2}.
The faster-is-slower effect was well explained as a consequence of the presence of human blocking clusters 
that obstruct the exit during an emergency evacuation \cite{Dorso1}.  \\ 

The SFM, in its basic version, was reported to be suitable for describing, 
at least qualitatively, a variety of emergency situations, including the presence
of obstacles, or the existence of more than one exit \cite{Dorso3,Dorso5}.
Scenarios described in Refs.~\cite{Cornes1,Dorso4,Dorso6} required, however, a step up implementation
although sustaining the basic model and its parameters.  \\

In the last years, several extensions of the SFM have been 
proposed. These extensions solve numerical pitfalls \cite{koster1} 
and other issues such as oscillations, overlapping and non-realistic trajectories
\cite{chraibi1,dietrich1}. However, many of these extensions do not correspond to the 
assumptions that motivated the original SFM~\cite{Helbing1,Helbing4}, focusing on the 
individual behavior of each agent. The basic SFM assumes strong interactions
in a dense environment.\\

Some questioning arose on the true psychological tendency of the pedestrians to 
stay away from each other. While the social forces accomplish this tendency, it 
attains a somewhat unrealistic ``colliding behavior'' for slowly moving 
pedestrians \cite{Lakoba}. His (her) repulsive tendency is expected to decrease 
as approaching a more crowded environment. The small fall-off length $B=0.08\,$m 
suggested by Helbing in Ref.~\cite{Helbing1} does not completely solve this 
issue. It neither agrees with the fact that pedestrians prefer to keep a 
comfortable $0.5\,$m distance between each other in a moderately crowded 
environment, nor it fits accurately the empirical velocities reported for 
crowds under normal conditions \cite{Lakoba}.  \\

Researchers turned back to examine the available data on the velocity and flux 
behavior for different density environments \cite{Boltes,helbing3,seyfried1,seyfried}. 
Ref.~\cite{helbing3} summarizes empirical data from the
literature, and their own data set, acquired from videos of the Muslim 
pilgrimage in Mina-Makkah (2006). They showed from the empirical fundamental 
diagram (flux $J$ versus density $\rho$) that highly dense crowds (seemingly up 
to $10\,$p~m$^{-2}$) do not drive the pedestrians velocity to zero, although the 
reasons for this remain rather obscure.   \\

More recent findings on the fundamental diagram for an extremely dense
 situation show that, sometimes, the flux may increase with the density (see 
Ref.~\cite{lohner1}). This seems to contradict the results of 
Ref.~\cite{helbing3}. But both results appear not to be completely  comparable 
since data was acquired at different points, and thus,
at different stages of the ritual. Ref.~\cite{lohner1}, indeed, states that the fundamental 
diagram may strongly depend on the type of motion and the nature of the event.\\

Our work focuses on data from Ref.~\cite{helbing3}. 
Further measurements can be found in 
Refs.~\cite{Boltes,lohner1,Dridi1,Dridi2,Baqui1}.
The PedFlow model developed by L\"ohner \textit{et al}. (Ref.~\cite{lohner2})  
is an alternative to the Helbing model. Togashi \textit{et al}. improved the 
calibration of this model using a Kalman filter ensemble \cite{Togashi1}. \\

The high density regime appears to be the most complicated one. Caution was 
claimed when (automatically) transferring the usual ``calibrated'' parameters 
of the SFM to this regime. It was argued that the pedestrians' body size 
distribution and the ``situational context'' are somewhat responsible for the 
unexpected departure from the original SFM parameters \cite{kwak,johansson1}. But other 
researchers pointed out that this departure actually expresses the lack of a 
mechanism to properly handle the pedestrians' ``required space to move''. Some 
modifications to the basic SFM were then proposed to overcome this difficulty 
\cite{parisi2,seyfried2}. \\

A mechanism allowing an ``increase of the space to move'' (particularly in relaxed/low
 density situations) is a compelling necessity in the context of the SFM.
But a sharp ``re-calibration'' of the model for high density situations appears not 
to be completely satisfactory \cite{johansson}. A more ``natural'' way of 
handling this matter requires a deep examination of the current SFM parameters. 
The net-time headway (roughly, the relaxation time) was first examined in 
Ref.~\cite{johansson}. The author sustains the hypothesis that the 
pedestrians net-time headway should increase until there is ``enough space to 
make a step''. He shows that a density dependent net-time headway is a 
suitable parameter to smartly reproduce the empirical fundamental diagram for 
highly dense crowds \cite{johansson}.  \\ 

Our own examination of the SFM parameters suggests that not only the net-time 
headway, but the friction between pedestrians (and with the walls) can 
reproduce the pattern of the fundamental diagram. Our working hypothesis is 
that friction is the crucial parameter in the dynamics of highly dense crowds. 
We actually sustain the SFM model with no further ``re-calibrations'', but 
with the right friction value, in order to meet the fundamental diagram 
pattern. \\  

We want to emphasize that although the friction value appearing in 
Ref.~\cite{Helbing1} is a commonly accepted estimate throughout the  
literature, other values have also been proposed \cite{colombi2017}. We 
propose our value as an experimentally based parameter, suitable for 
highly dense crowds. \\

The investigation is organized as follows. We first recall the SFM in 
Section~\ref{sfm}, while including the precise definitions for flux, density 
and clustered structures. Section~\ref{simulations} presents our numerical 
simulations for pedestrians moving along corridors. The hypotheses of the 
work are stated in Section~\ref{Hypotheses}. The corresponding results 
are shown in Section~\ref{results}. Our main conclusions are detailed in Section~\ref{conclusions}.\\     

\section{\label{background}Background}

\subsection{\label{sfm}The Social Force Model}

Our research was carried out in the context of the ``Social Force Model'' (SFM) 
proposed by Helbing \textit{et al.} \cite{Helbing1}. This 
model states that human motion is caused by the desire of people to reach a 
certain destination at a certain velocity, as well as other environmental factors. The pedestrians' 
behavioral pattern in a crowded environment can be modeled by three kinds of 
forces: the ``desire force'', the ``social force'' and the ``granular force''.\\

The ``desire force'' represents the pedestrian's own desire to reach a 
specific target position at a desired velocity $v_d$. But, in order to reach 
the desired target, he (she) needs to accelerate (decelerate) from his (her) 
current velocity $\mathbf{v}^{(i)}(t)$. This acceleration (or deceleration) 
represents a ``desire force'' since it is motivated by his (her) own 
willingness. The corresponding expression for this forces is 

\begin{equation}
        \mathbf{f}_d^ {(i)}(t) =  
m_i\,\displaystyle\frac{v_d^{(i)}\,\mathbf{e}_d^
{(i)}(t)-\mathbf{v}^{(i)}(t)}{\tau}, \label{desired}
\end{equation}

\noindent where $m_i$ is the mass of the pedestrian $i$. $\mathbf{e}_d$ 
corresponds to the unit vector pointing to the target position and $\tau$ is a 
constant related to the relaxation time needed to reach his (her) desired 
velocity. For simplicity, we assume that $v_d$ remains constant during the 
entire process and is the same for all individuals, but $\mathbf{e}_d$ changes 
according to the current position of the pedestrian. Detailed values for $m_i$ 
and $\tau$ can be found in Refs.~\cite{Helbing1,Dorso3}. \\

The ``social force'' represents the psychological tendency of any two pedestrians,  
say $i$ and $j$, to stay away from each other. It is represented by a repulsive interaction force 

\begin{equation}
        \mathbf{f}_s^{(ij)} = A_i\,e^{(R_{ij}-r_{ij})/B_i}\mathbf{n}_{ij}, 
        \label{social}
\end{equation}

\noindent  where $(ij)$ means any pedestrian-pedestrian pair, or pedestrian-wall 
pair. $A_i$ and $B_i$ are fixed values, $r_{ij}$ is the distance between  the 
center of mass of the pedestrians $i$ and $j$, and the distance $R_{ij}=R_i+R_j$ 
is the sum of the pedestrians radius. $\mathbf{n}_{ij}$ means the unit vector in 
the $\vec{ji}$ direction.\\

Any two pedestrians touch each other if their distance $r_{ij}$ is smaller than 
$R_{ij}$. Analogously, any pedestrian touches a wall if his (her) distance $r_{ij}$ to the wall is smaller than $R_i$. In these cases, an additional force is included in the model, called the ``granular force''(\textit{i.e.} friction force). This force is considered to be a linear function of the relative (tangential) velocities of the contacting individuals. In the case of the friction exerted by the wall, the force is a linear function of the pedestrian tangential velocity. 
Its mathematical expression reads

\begin{equation}
        \mathbf{f}_g^{(ij)} = 
\kappa\,(R_{ij}-r_{ij})\,\Theta(R_{ij}-r_{ij})\,\Delta
\mathbf{v}^{(ij)}\cdot\mathbf{t}_{ij}, 
        \label{granular}
\end{equation}

\noindent where $\kappa$ is the friction coefficient. The function 
$\Theta(R_{ij}-r_{ij})$ is zero when its argument is negative (that is, 
$R_{ij}<r_{ij}$) and equals unity for any other case (Heaviside function). 
$\Delta\mathbf{v}^{(ij)}\cdot\mathbf{t}_{ij}$ represents the difference between 
the tangential velocities of the sliding bodies (or between the individual and 
the walls).\\

The above forces actuate on the pedestrians dynamics by changing his (her) 
current velocity. The equation of motion for pedestrian $i$ reads

\begin{equation}
m_i\,\displaystyle\frac{d\mathbf{v}^{(i)}}{dt}=\mathbf{f}_d^{(i)}
+\displaystyle\sum_{j=1}^{N}\displaystyle\mathbf{f}_s^{(ij)}
+\displaystyle\sum_ {
j=1}^{N}\mathbf{f}_g^{(ij)},\label{eq_mov}
\end{equation}

\noindent where the subscript $j$ represents all the other pedestrians or walls
(excluding pedestrian $i$).\\

In the original model, there is no distinction between the friction coefficient of pedestrian-pedestrian interaction and pedestrian-wall interaction. Both interactions are modeled with the same constant estimated parameter $\kappa$. In this paper, we analyze situations in which the friction coefficient may take different values. We define $\kappa_i$ and $\kappa_w$ as the friction coefficient related to the pedestrian-pedestrian interaction and the pedestrian-wall interaction, respectively. 

\subsection{\label{fundamental-diagram} Fundamental Diagram}

Inspired from vehicular traffic dynamic studies, many researches on pedestrian 
dynamics focus their attention on the relation between the flow and the density 
of a moving crowd. This relation is represented by the ``fundamental diagram" and it 
has become one of the most common ways to characterize the pedestrians' dynamics 
along a corridor in unidirectional and bidirectional flows 
\cite{seyfried1,fruin1,mori1,polus1,jelic1}. \\

We follow the same definition as in Ref.~\cite{helbing3} regarding the fundamental diagram analysis. That is, the local density at place $\vec{r}=(x,y)$ and time $t$ is given by the following expression

\begin{equation}
\rho(\vec{r},t)=\sum_{j}f(\vec{r}_j(t)-\vec{r}), \label{ec-density}
\end{equation}

\noindent where function $f(\vec{r}_j(t)-\vec{r})$ is a Gaussian distance-dependent weight function defined as

\begin{equation}
f(\vec{r}_j-\vec{r})=\frac{1}{\pi R^2}\exp[-\left \| \vec{r}_j-\vec{r} \right \|^2/R^2]. \label{ec-f}
\end{equation}

\begin{figure}[htbp!]
\centering
\includegraphics[width=0.7\columnwidth]
{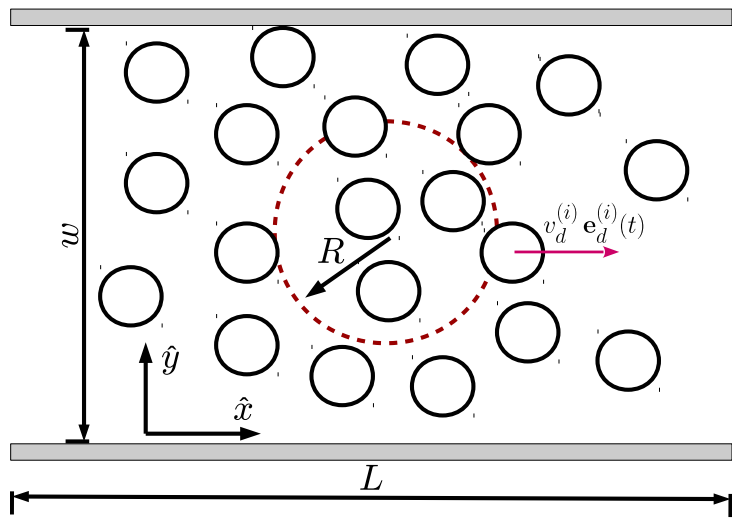}
\caption{\label{corridor} Schematic diagram for individuals in the corridor. 
The circles represent pedestrians moving from left to right. $w$ represents the 
corridor width, $L$ represents the length. The rectangular boxes are upper and 
lower blocks that represent the walls of the corridor. The dashed circle in the 
middle corresponds to the fundamental diagram measurement region.}
\end{figure}

\noindent $\vec{r}_j(t)$ is the position of the pedestrians $j$ in the surroundings of $\vec{r}$ and $R$ is a parameter that ponders more significantly to the pedestrians inside the circle shown in Fig.~\ref{corridor} . 
The local speeds are defined as the weighted average  

\begin{equation}
\vec{V}(\vec{r},t)=\frac{\sum_j \vec{v}_jf(\vec{r}_j(t)-\vec{r}) }{\sum_j f(\vec{r}_j(t)-\vec{r})}, \label{ec-v}
\end{equation}

\noindent while flow is determined according to the fluid-dynamic formula

\begin{equation}
\vec{J}(\vec{r},t)=\rho(\vec{r},t)\vec{V}(\vec{r},t). \label{ec-flow}
\end{equation}

It is well known that the original version of the SFM  is incapable of reproducing the 
fundamental diagram at high densities (say above 
5~p~m$^{-2}$) \cite{parisi2} . Different 
approaches were proposed to fix this drawback: increasing the net-time 
headway~\cite{johansson}, canceling the desired velocity~\cite{parisi2} or even 
inducing the jamming state by an attraction target~\cite{kwak}. These approaches 
seem suitable when individuals are not so anxious to reach a certain 
destination. However, when individuals are escaping or rushing due to a 
stressful situation, pedestrians would neither consider reducing their desired 
force nor being captured by an attraction. \\

\subsection{\label{granular-cluster} Clustering structures}

A characteristic feature of pedestrian dynamics is the formation of clusters. Clusters of pedestrians can 
be defined as the set of individuals that for any member of the group (say, $i$) there exists at least another member belonging to the same group ($j$) in contact with the former. Thus, we define a ``granular cluster'' ($C_g$) following the mathematical formula given in Ref.~\cite{Dorso1}

\begin{equation}
C_g:P_i~\epsilon~ C_g \Leftrightarrow \exists~ j~\epsilon~C_g / r_{ij} < R_i+R_j, \label{ec-cluster}
\end{equation}

\noindent where ($P_i$) indicates the \textit{ith} pedestrian and $R_i$ is his (her) radius 
(shoulder width). That means, $C_g$ is a set of pedestrians that interact not 
only with the social and the desired forces, but also with granular forces 
(\textit{i.e.} friction forces). The size of the cluster is defined as the 
number of pedestrians belonging to it. The fraction of clustered individuals is 
defined as the ratio between clustered individuals with respect to the total 
number of individuals in the crowd. In Section \ref{clusters} we analyze the 
clustering structures in terms of these two observables. \\

\section{\label{simulations}Setting and parameters}

We explored the flow of pedestrians along a straight corridor of length $L=$ 
28~m (with periodic boundary conditions) and different values of the width $w$. We explored 
widths ranging from $w=2$~m to $w=40$~m. The corridor had two side walls, placed 
at $y=0$ and $y=w$, respectively. The length of each wall was $L$. The 
pedestrians were modeled as soft spheres of radius $R_i=0.23$~m. This size was 
fixed according to Ref.~\cite{metric_handbook}. Initially, the individuals were 
randomly distributed along the corridor with a fixed global density 
(number of pedestrians over area) and with random initial velocities, resembling a 
Gaussian distribution with null mean value. We explored global density values in 
the range 1~p~m$^{-2}$~$<\rho<$~9~p~m$^{-2}$. 
We did not explore extreme densities (say $\rho>$9~p~m$^{-2}$) 
because we excluded injuring situations. The number of pedestrians in the 
simulation was given by the global density and the corridor dimensions chosen in 
each case. \\

In this work we use the common value $\kappa=2.4 \times 
10^{5}$ Kg~m$^{-1}$~s$^{-1}$, but we eventually set the newly defined 
parameters $\kappa_i=2.4 \times 10^{6}$ Kg~m$^{-1}$~s$^{-1}$  and $\kappa_w=2.4 
\times 10^{6}$ Kg~m$^{-1}$~s$^{-1}$, being $\kappa_i$ and $\kappa_w$ the 
pedestrian-pedestrian friction coefficient and the pedestrian-wall friction 
coefficient, respectively. \\

The desired velocity for each pedestrian $i$ was 
$\vec{v}_d^{~(i)}=1$~m~s$^{-1}$~$\hat{e}_d^{~(i)}$, where the target $\hat{e}_d^{~(i)}$ 
was set as $\hat{e}_d^{~(i)}=(L,y_i)\left \| (L,y_i) \right \|^{-1}$, being $L$ 
the \textit{x}-location of the end of the corridor and $y_i$ the 
\textit{y}-location corresponding to the \textit{ith} pedestrian 
(see Fig.~\ref{corridor}). This allowed the pedestrians to 
move from left to right in a unidirectional flow. Pedestrians that surpassed 
$x=L$ were re-injected at $x=0$, preserving their current velocity and 
\textit{y}-location (\textit{i.e.} periodic boundary conditions). This mechanism 
was carried out in order to keep
the crowd size unchanged.\\

We warn the reader that, for simplicity, we will not include the units 
corresponding to the numerical results. Remember that the friction coefficient 
has units $\left [ \kappa \right]=$Kg~m$^{-1}$~s$^{-1}$, the density $\left [ 
\rho \right]=$p~m$^{-2}$ and the flow $\left [ J \right 
]=$p~m$^{-1}$~s$^{-1}$.\\

\subsection{\label{software} Simulation software}

The simulations were implemented on LAMMPS molecular dynamics 
simulator with parallel computing capabilities \cite{plimpton}. The time 
integration algorithm followed the velocity Verlet scheme with a time step of 
$10^{-4}$~s. All the necessary parameters were set to the same values as in 
previous works (see Refs.~\cite{Dorso5,sticco}), except for the friction 
coefficient $\kappa$ and the radius $R_i$. \\

We implemented special modules in C++ for upgrading the LAMMPS 
capabilities to perform the SFM simulations. We also checked over 
the LAMMPS output with previous computations (see Refs. \cite{Dorso2, 
Dorso1,Dorso3, Dorso4,Dorso6}).\\

The measurements were taken once the system reached the 
stationary state ($t=30$~s), while the configurations of the systems were 
recorded every 0.05~s, that is, at intervals as short as 10\% of the 
pedestrian’s relaxation time (see Section. \ref{sfm}). The recorded magnitudes 
were the pedestrian’s positions and velocities for each process. We also 
computed the clustered structures using a LAMMPS built-in function 
named ``compute cluster-atom". This function assigns each pedestrian a cluster ID.
A cluster is defined as a set of pedestrians, each of which is within the cutoff
distance from one or more other pedestrians in the cluster. 
The cutoff was set as $R_i+R_j$ to assimilate the LAMMPS built-in function with the 
cluster definition given in  Eq.~(\ref{ec-cluster}).
If a pedestrian has no neighbors within the cutoff distance, then it is a 1-pedestrian cluster.\\

\section{\label{Hypotheses} Hypotheses}

We stress the fact that our investigation focuses on moving
pedestrians in a high density situation (say, the one experienced at
the Jamaraat bridge). As mentioned in Section \ref{introduction}, a deep
examination of the (basic) SFM parameters is required before
proceeding to any extension of the model.\\

Recall from the video analysis of the Muslim pilgrimage in Mina/Makkah
(see Section \ref{introduction}) that high-density flows can attain zero velocity, as people
start pushing to gain space \cite{helbing3}. This unexpected behavior can not be
reproduced by the (basic) SFM, it might happen because of ``an
underestimation of the local interactions triggered by high densities"
\cite{yu1}, or, the absence of a ``delayed reaction in cases of unexpected
behaviors" \cite{johansson}. Both statements are currently working hypotheses since
experimental data (specifically, measurements of pedestrian flux and
densities) do not ``provide any insight into the mechanisms and
dynamics behind the pedestrians interactions and behaviors" \cite{johansson}.\\

Researchers propose a ``re-calibration" of the (basic) SFM, in order to
attain ``stop-and-go" flows for highly dense crowds
\cite{johansson,yu1}. Presumably, this kind of instabilities within the crowd prevent
people from stopping at extremely high densities. The intended
``re-calibration" consists of either enhancing the (local) social
interactions or increasing the net-time headway (roughly, the
relaxation time) for the high density regime. Both extensions,
however, may not exclude other possibilities involving not only
individual motion but collective (mass) motion \cite{helbing3}. Researchers
further point out that the relevance of physical contact in extremely
dense crowds may suppose a somewhat commonality with granular
media exists \cite{helbing3}.\\

We show that the experimental data (say, the flux-density diagram) can be 
modeled under quasi-stationary conditions in the high density regime.\\

Our starting point will be the re-examination of the (basic) SFM. 
We propose a reduce-in-units equation of motion of the SFM (see Section~\ref{appendix_1} 
for details). From this point of view, the parameter $A$ standing for the intensity of the
 social force $\mathbf{f}_s$ is replaced by the reduced-in-units parameter $\mathcal{A}$.
 The friction intensity $\kappa$ of the granular force $\mathbf{f}_g$ is replaced by the reduced-in-units
 parameter $\mathcal{K}$. No other parameters are required in the reduced-in-units model
 since $m$, $v_d$, $\tau$ and $B$ are all included in $\mathcal{A}$ and $\mathcal{K}$.
 The desired force $\mathbf{f}_d$, indeed, will only depend on the target direction. In Section~\ref{appendix_1} we derivate the 
reduce-in-units equation of motion and the meaning of the control parameters ($\mathcal{A}$ and $\mathcal{K}$). \\

 We presume that physical contact is
a key feature in dense crowds, despite the fact that other issues may
also contribute to the flow reduction \cite{johansson1}. However, the latter could
be satisfactorily omitted in past research \cite{johansson}, and thus, we will not
attempt to introduce further extensions to the (basic) SFM for the
sake of simplicity.\\

The former ``re-calibrations" accomplish the socio-psychological
response of the crowd to ``gain more space" (by either enhancing local
interactions or performing a delayed reaction). We are aware that
crowds may respond differently in many situations (see Ref.~\cite{drury1}). Our working
hypothesis, however, does not focus on the crowd socio-psychological
response, but on the physical contact among pedestrians (and the
walls). The socio-psychological attitude of the pedestrians will be
assumed to remain fixed along the simulations (with the 
desired speed limited to $v_d=1\,$ m~s$^{-1}$). \\

Our investigation appears somewhat restricted to the (almost)
unidirectional flow inspired from the Muslim pilgrimage in Mina/Makkah.
This means that the following ``re-calibration" results hold for
corridor-like situations, and are not intended to be (automatically)
translated to bottleneck situations. Neither can be extended to other
boundary conditions (say, no limiting walls) since the boundary is a
key feature of collective motion. Nevertheless, our results accomplish
the available data on the Hajj pilgrimages \cite{helbing3,lohner1}.

\section{\label{results}Results}

\subsection{\label{fundamental_diagram} Fundamental diagram in the original model}

In this Section we present the results relating the local flow, velocity and density (\textit{i.e.} the fundamental diagram). The measurements were taken in the middle of the corridor using the definitions given in Eq.~(\ref{ec-v}), Eq.~(\ref{ec-flow}), and as shown in Fig.~\ref{corridor}. All the results shown here correspond to $R=1$~m (see Eq.~(\ref{ec-flow}) and Fig.~\ref{corridor}). We further varied $R$ until $R=3$, but no significant changes were observed. \\

\begin{figure}[htbp!]
\centering
\includegraphics[width=0.7\columnwidth]
{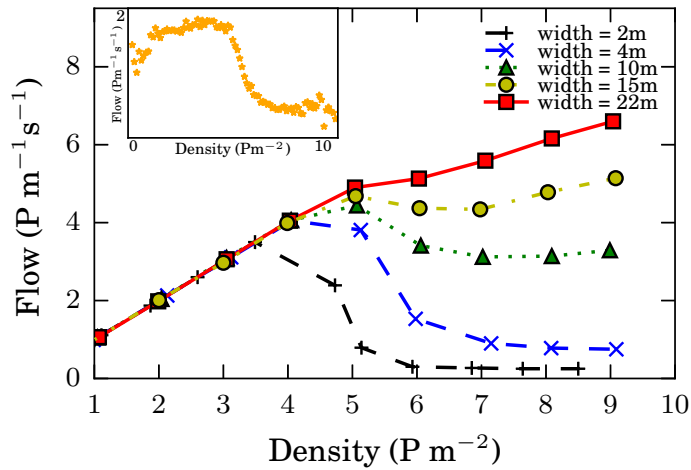}
\caption{\label{fundamental_diagram_flow} Mean flow ($J$) as a function of the 
density ($\rho$) for different widths. Initially, pedestrians were randomly 
distributed along the corridor. The measurements were taken in the middle
of the corridor once the system reached the stationary state (see Fig.~\ref{corridor}). The length of the corridor 
was $L=$28~m for all cases (with periodic boundary conditions in the \textit{x} direction). The inset corresponds to Ref.~\cite{helbing3}}
\end{figure}

Fig~\ref{fundamental_diagram_flow}, shows the fundamental diagram (flow vs. density) for different corridor widths. We can distinguish the two typical regimes of the fundamental diagram. In the free flow regime ($\rho < 5$), the flow increases linearly with the density since collisions between pedestrians (and with the walls) are scarce. Pedestrians are able to achieve their desired velocity, leading to a flow that grows linearly with the density ($J \propto \rho$) until $\rho=5$. This behavior applies to all the analyzed corridor widths.\\

On the other hand, we have the congested branch  for $\rho > 5$. Here we face two different scenarios:

\begin{itemize}
\item[(i)] For narrow corridors (say $w < 10$) we can see that the flow reduces as the density increases. This resembles the traditional behavior of the fundamental diagram reported in the literature. 
\item[(ii)] For wide corridors (say $w > 15$) we see that the flow increases with density. This contradicts the typical behavior of the fundamental diagram.   
\end{itemize}

In the case of narrow corridors, both the simulated case and the empirical results converge to a constant flow value. It is remarkable that the system does not reach a freezing state such as the one reported in Refs.~\cite{kwak,lin}. Recall that our simulations do not include any respect factor (see Ref.~\cite{parisi2}), or changes in the net-time headway (Ref.~\cite{helbing3}), or the urge to see an attraction (Ref.~\cite{kwak}). We assume a well-defined target and the same $v_d$ for all the pedestrians.\\

The inset in Fig.~\ref{fundamental_diagram_flow} corresponds to the empirical data from Helbing (see Ref.~\cite{helbing3}) at the entrance of the Jamaraat bridge (the corridor width was $w=22$~m). Notice that our results from simulations corresponding to a $w=22$~m corridor, exhibit a different behavior along the congested regime. In the results from simulations, the flow increases even for the highest explored density. On the contrary, the empirical data exhibit a flow reduction for $\rho > 5$ until reaching a plateau for the highest explored density values.  \\

In order to fulfill the experimental fundamental diagram, it becomes necessary that the flow at the maximum explored density ($\rho_{max} = 9$) does not exceed the flow at $\rho = 5$ (upper bound). That is:  $J(\rho = 9) < J(\rho = 5)$. From the flow definition in Eq.~(\ref{ec-flow}) we can derive the bounding values

\begin{equation}
v(\rho_{max}) < \frac{5v_d}{\rho_{max}} \leq \frac{5}{9} v_d,\\
\end{equation}

As our desired velocity is fixed at $v_d = 1$~m~s$^{-1}$, we conclude that the speed at 
the maximum density has to be bounded by $v(\rho_{max}) \lesssim  0.5$~m~s$^{-1}$ in 
order to satisfy the qualitative behavior of the (experimental) fundamental 
diagram reported in the literature.\\

\begin{figure}[htbp!]
\centering
\includegraphics[width=0.7\columnwidth]
{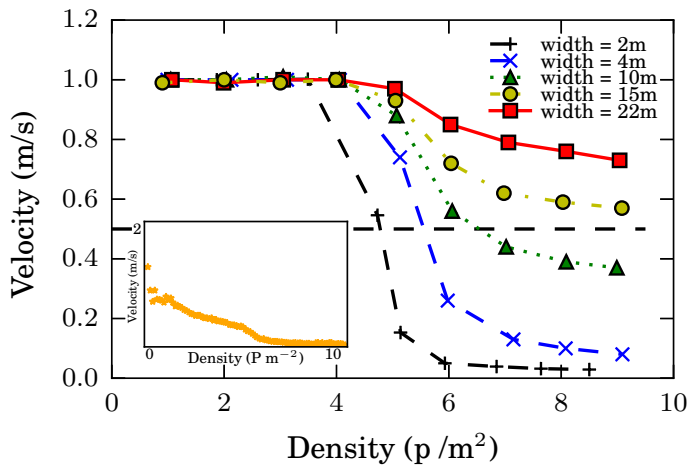}
\caption{\label{fundamental_diagram_speed} Mean speed ($V$) as a function of the density ($\rho$) for different widths. Initially, 
pedestrians were randomly distributed along the corridor. The measurements were taken in the middle
of the corridor once the system reached the stationary state (see Fig.~\ref{corridor}). The length of the corridor 
was 28~m in all cases (with periodic boundary conditions in the \textit{x}-direction). The inset corresponds to Ref.~\cite{helbing3}.}
\end{figure}

The above reasoning is consistent with the speed-density results shown in Fig.~\ref{fundamental_diagram_speed}. As a visual guide, we plotted $v=0.5$~m~s$^{-1}$ with a horizontal dashed line. The close examination of $\rho_{max} = 9$ shows that values corresponding to the wide corridors ($w=15$~m and $w=22$~m) exceed $v=0.5$~m~s$^{-1}$. But, those values corresponding to narrow corridors fall below $v=0.5$~m~s$^{-1}$. \\

The results shown in Fig.~\ref{fundamental_diagram_speed} confirm the fact that when the density is low enough, pedestrians manage to walk at the desired velocity ($v=v_d=1$~m~s$^{-1}$). Above $\rho>5$, however, the velocity begins to slow down. The inset shows the experimental data at the entrance of the Jamaraat bridge. We may conclude that our simulations agree with the experimental data for narrow corridors, but disagree as these become wider. The wider the corridor, the greater the velocity for all the density values explored. In Section \ref{velocity_profile} we will further discuss this topic.\\

It should be pointed out that the Jamaraat data do not exhibit a ``really" constant velocity for low densities. But this seems reasonable since our simulations do not include the complexities of the real situation when the density is low. We will not analyze this phenomenon in this investigation. \\

We may summarize our first results as follows. We were able to validate the 
original SFM for narrow corridors through the fundamental 
diagram. However, the SFM (in its current version) disagrees with experimental 
data as the corridors widen. We will focus in the next Section on the velocity 
profile in order to investigate this discrepancy. 

\subsection{\label{velocity_profile} Velocity profile}

As we mentioned in Section \ref{fundamental_diagram}, when the density is low, pedestrians achieve the desired velocity ($v=v_d=1$~m~s$^{-1}$). Since the results of the previous Section only hold for the area located in the middle of the corridor (see the dashed circle in Fig.~\ref{corridor}), we want to shed some light and understand what is happening across the entire corridor.\\

We first noticed that low-density situations ($\rho<5$) lead to a cruising velocity profile $v=v_d$. This is valid for every location in the corridor (not only the center as was previously noticed in Section \ref{fundamental_diagram}). For higher densities ($\rho>5$), the velocity profile turns into a parabola-like function. This shape resembles the usual velocity profile for laminar flow in a viscous fluid, where the velocity increases toward the center of a tube. In our case, pedestrians near the walls are the ones with the lower velocity. The velocity increases when departing from the wall until it reaches the maximum at the center of the corridor. This behavior suggests that the wall friction on the pedestrians, is playing a relevant role in the velocity distribution. The velocity profile is in agreement with the empirical data in Ref.~\cite{zhang1}.  \\

\begin{figure}[htbp!]
\centering
\includegraphics[width=0.7\columnwidth]
{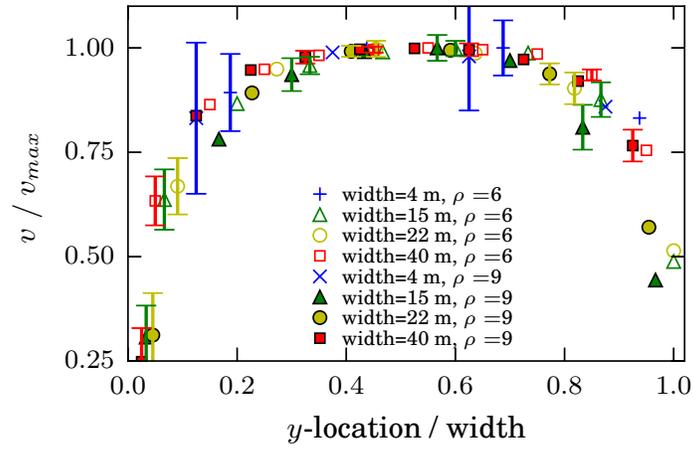}
\caption{\label{speed-profile-width-normaliz} Scaled velocity profile (normalized velocity) vs. $y$-location for different corridors width (see legend for the corresponding widths) and two different densities. Empty markers correspond to $\rho=6$ while filled markers correspond to  $\rho=9$. The simulated corridor was 28~m length. Pedestrians walk from left to right with periodic boundary condition in the $x$-direction. Initially, pedestrians were randomly distributed. The horizontal axis is normalized by the corridor width, the vertical axis is normalized by the maximum velocity reached in each case.  The bin size was 1~m for all cases except for $w=4$~m since the bin was 0.5~m.}
\end{figure}

Fig.~\ref{speed-profile-width-normaliz} exhibits the scaled velocity profile. The horizontal axis is normalized by the corresponding corridor width. The vertical axis is normalized by the maximum velocity ($v_{max}$) corresponding to each data set. Filled markers correspond to density $\rho=9$, while empty markers correspond to $\rho=6$. Notice that all the data follow the same  pattern, suggesting that the velocity profile exhibits a somewhat fundamental behavior, regardless of the scale of the corridor (and the density). Hence, the velocity growth rate from the wall towards the center of the corridor is the same in spite of the size of the corridor width and the density. \\

We remark that there is a clear relation between $v_{max}$ and the corridor width. That is, the wider the corridor, the higher the maximum attained velocity (Fig.~\ref{speed-profile-width-normaliz} does not exhibit this behavior because the velocity is normalized by $v_{max}$).\\

In summary, the scaled velocity profile (see Fig.~\ref{speed-profile-width-normaliz}) does not report any relevant difference as the corridor widens (withing the high density regime). This suggests that the pedestrian dynamics remain essentially the same. The maximum attainable velocity ($v_{max}$), however, seems to be a sensible parameter with respect to the flux.  The narrow corridors attain lower values of $v_{max}$ and thus lower flux. We may expect the flow not to increase if $v_{max}$ remains low enough along the explored density range.\\

\subsection{Work done by friction force}

In the previous Section we studied the velocity profiles for different corridors as a function of density. Here we present the spatial distribution of the work done by the friction force. The pedestrian-pedestrian friction and the pedestrian-wall friction were computed.
The work on each pedestrian $i$ was numerically obtained through the integration Trapezoidal rule (according to Eq.~\ref{ec-trapezoidal}). The integration time step was $\Delta t = 0.05$~s. 

\begin{equation}
W^{(i)}(t) \simeq \left [ \vec{f}^{(i)}(t+\Delta t) + \vec{f}^{(i)}(t)  \right ]\cdot \frac{\Delta \vec{x}}{2}. \label{ec-trapezoidal}
\end{equation}

\begin{figure}[!htbp]
\centering
    \subfloat[]{\includegraphics[width=0.5\columnwidth]{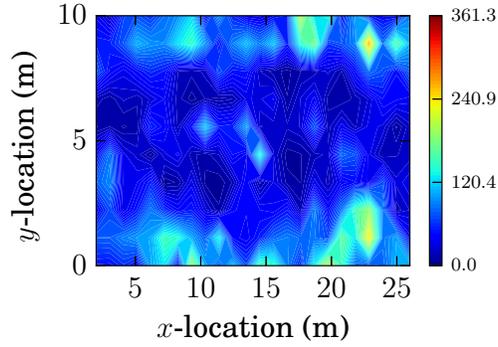}\label{abswg_width10}}\\ 
    \subfloat[]{\includegraphics[width=0.5\columnwidth]{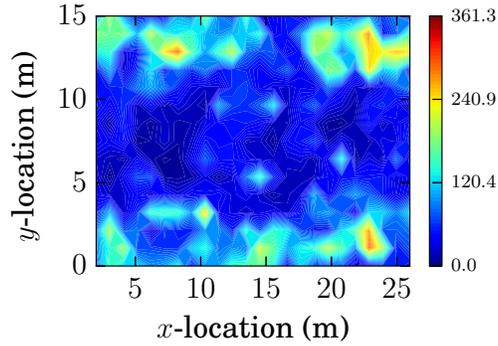}\label{abswg_width15}}\\
    \subfloat[]{\includegraphics[width=0.5\columnwidth]{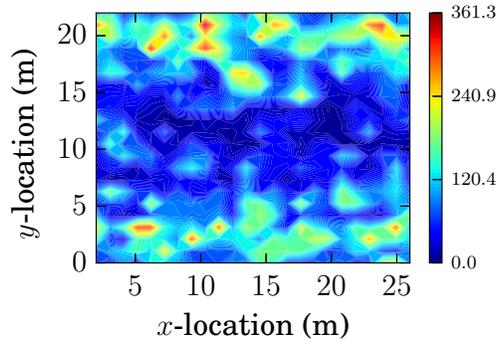}\label{abswg_width22}}\
\caption[width=0.47\columnwidth] {Color map for the absolute value of the work done by the friction. The density chose was $\rho=6$. (a) Corresponds to $w=$10~m (b) $w=$15~m and (c) $w=$22~m. The axis represent the location in the corridor ($x$ and $y$). The scale bar on the right is expressed in Joule units. The work was numerically integrated following the Trapezoidal rule with $\Delta t =0.05$~s. The pedestrians desired velocity was $v_d = 1$~m~s$^{-1}$. The contour lines were computed on a square grid of 1m$\times$~1m and then splined to get smoother curves.} 
\label{abswg}
\end{figure}

Once the work on every pedestrian is calculated, we proceeded to bin the corridor into a squared grid of 1~m$\times$1~m cells in order to associate the work values with the corresponding spatial location. Fig.~\ref{abswg} shows three color maps of the absolute value of the work done by the friction force. The horizontal and vertical axis represent the $x$-location and $y$-location of the corridor respectively. The color map associates higher work values with red colors and lower work for blue colors. The walls are located at $y=0$ and $y=w$ (bottom and top of each figure). Fig.~\ref{abswg_width10} corresponds to a 10 m width corridor, Fig.~\ref{abswg_width15} corresponds to a 15 m width corridor and Fig.~\ref{abswg_width22} corresponds to a 22 m width corridor.\\

In the three figures, we observe a similar pattern: the regions near the walls (bottom and top) are the most dissipative ones. The center of the corridor is though not a very dissipative region. Furthermore, the work seems to increase with the corridor width. This occurs because the relative velocity between pedestrians is greater in the wide corridors than in the narrow ones. The wider the corridor, the greater the slope (corresponding to locations near the wall).\\

Recall from Eq.~(\ref{granular}), that the friction force depends on the 
compression and the relative velocity among pedestrians. The compression levels 
remain the same in the three cases since the compression only depends on the 
density (which is fix at $\rho=6$ for the three color maps). Thus, the 
differences between Figs.~\ref{abswg_width10}, \ref{abswg_width15} and 
\ref{abswg_width22} can only be explained by the increment of the relative velocity 
between individuals. \\

The above observations drive the following conclusion. The maximum attainable velocity $v_{max}$ accomplishes a maximum velocity slope (with its associated friction dissipation). Both the friction with the walls and the friction between the pedestrians appear as relevant magnitudes for properly slowing down the crowd velocity, in order to fit the experimental data. Section \ref{appendix_2} supports this assertion with a simple example.

\subsection{\label{appendix_1}The reduced equation of motion}

The equation of motion within the context of the SFM includes at least six parameters ($m$, $\tau$, $A$, $B$, 
$\kappa$ and $v_d$), but the equation itself barely depends on two. The process 
of parameter's reduction is achieved by defining the (reduced) magnitudes 

\begin{equation}
 \left\{\begin{array}{lll}
         t' & = & t/\tau \\
         r' & =& r/B \\
         v' & = & v/v_d \\
        \end{array}\right.
\end{equation}

The (reduced) equation of motion reads

\begin{equation}
\displaystyle\frac{d\mathbf{v}'}{dt'}=
\displaystyle\frac{\tau}{m\,v_d}\bigg(\mathbf{f}_d+
\mathbf{f}_s+\mathbf{f}_g\bigg).\label{eqn_appendix_1}
\end{equation}

It is straight forward from Eq.~(\ref{eqn_appendix_1}) that the 
corresponding reduced forces can 
be defined as follows\\

\begin{equation}
 \left\{\begin{array}{lll}
         \mathbf{f}_d' & = & \hat{\mathbf{e}}_d-\mathbf{v}' \\
         && \\
         \mathbf{f}_s' & =& \mathcal{A}\,\exp(r'-d')\,\hat{\mathbf{n}} \\
         && \\
         \mathbf{f}_g' & = & 
\mathcal{K}\,(2r'-d')\,\Theta(2r'-d')\,(\Delta\mathbf{v}'\cdot\hat{\mathbf{t}})\
 
\hat{\mathbf{t}} \\
        \end{array}\right.\label{eqn_appendix_2}
\end{equation}

\noindent where $\mathcal{A}=A\tau/(m\,v_d)$ and 
$\mathcal{K}=\kappa B\tau/m$. \\ 

Notice that $\mathcal{A}$ and $\mathcal{K}$ are actually the only two 
control parameters in Eq.~(\ref{eqn_appendix_1}) for identical 
pedestrians. The ratio $\tau/m$ is common to both, but the magnitudes 
$Av_d^{-1}$ and $\kappa B$ handle each parameter separately.\\

We envisage $\mathcal{A}$ as a parameter related exclusively to the 
repulsion between pedestrians and $\mathcal{K}$  as a parameter related to the friction-repulsion.
While the former is valid in every situation, the latter only takes part when the
pedestrians are in contact.\\

The fact that $\mathcal{A}$ and $\mathcal{K}$ share the parameter $\tau$ is 
in agreement with the conclusions outlined in Ref.~\cite{johansson}. The 
relaxation time (or ``net-time headway'') $\tau$ actually ``weights'' the 
effects of the environment on the individual (that is, the social repulsion and 
the friction), and thus, appears as a ``key control parameter'' for the 
fundamental diagram as claimed in Ref.~\cite{johansson}.\\

The role of $\tau$ may be somewhat ambiguous whenever the social 
repulsion becomes negligible with respect to the friction. This may occur if 
some kind of balance exists between neighboring pedestrians in 
symmetrical configurations (\textit{i.e.} in crowded corridors). We may 
hypothesize that the ``key control parameter'' may correspond to either 
$\tau$, or, the friction itself $\kappa$. This is an open question, and a 
first order approach to this matter is outlined in Section~\ref{appendix_2}.\\

\subsection{\label{appendix_2}A simple model for the corridor}

\begin{figure}[htbp!]
\centering
\includegraphics[width=0.7\columnwidth]
{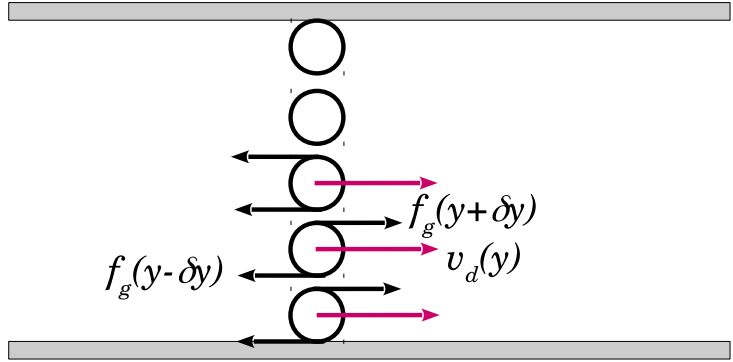}
\caption{\label{pasillo} Schematic diagram for individuals in a corridor. 
The circles represent pedestrians moving from left to right. The desired force 
(red arrows) and sliding friction (black arrows) are assumed to be the only 
relevant forces.}
\end{figure}

A toy model for a moving crowd along a corridor is the one represented 
schematically in Fig.~\ref{pasillo}. Pedestrians (circles in 
Fig.~\ref{pasillo}) 
are assumed to be lined up from side to side across the corridor, at any given 
position. Social forces in the $x$-direction are further considered to vanish 
because of the translational symmetry. Thus, only the sliding friction is allowed 
to balance the pedestrians own desire. The (reduced) movement equation for 
the $x$-direction according to Section \ref{appendix_1} and Fig.~\ref{pasillo} 
is  

\begin{equation}
\displaystyle\frac{dv'}{dt'}(y')=1-v'(y')+f_g'(y'+\delta 
y')-f_g'(y'-\delta y'),\label{eqn_appendix_2:1}
 \end{equation}

where $v'(y')$ corresponds to the (reduced) velocity (for the $x$-direction) of 
the individual located at the $y'$ position. Notice that the individuals 
remain at the same $y'$ position while traveling through the corridor since 
balance is expected to take place across the corridor. These positions are 
roughly $\delta y'$, $3.\delta y'$, $5.\delta y'$,.... Actually, it is not 
relevant (for now) the value of $y'$, and a further simplification can be done 
by labeling $v'(y')=v_i$ and $v'(y'\pm 2.\delta y')=v_{i\pm 1}$. The velocity 
of the individual in contact with the bottom wall in Fig.~\ref{pasillo} will be 
labeled as $v_1$. \\

The last two terms in Eq.~(\ref{eqn_appendix_2:1}) correspond to the 
net drag applied on the pedestrian with velocity $v_i$. According to 
Eq.~(\ref{eqn_appendix_2}) this drag may be expressed as

\begin{equation}
 f_{g,i+\frac{1}{2}}'-f_{g,i-\frac{1}{2}}'=
  \left\{\begin{array}{lcl}
          2\alpha\,v_{2}-3\alpha\,v_{1} &  & i=1 \\
          & & \\
          2\alpha\,(v_{i+1}-2v_i+v_{i-1}) & & i>1\\
         \end{array}\right.\label{eqn_appendix_2:2}
\end{equation}

\noindent for $\alpha=\mathcal{K}(r'-\delta y')$. Recall that our first order 
approach considers $\delta y'$ as roughly uniform across the corridor. \\

The stationary situation can be computed straight forward from 
Eq.~(\ref{eqn_appendix_2:1}). Thus, for $\dot{v}_{i}=0$ the following set of 
equations determine the velocity profile in the corridor (within this toy 
model)

\begin{equation}
  \left\{\begin{array}{lcl}
          (3\alpha+1)\,v_{1} - 2\alpha\,v_{2} & = & 1 \\
          & & \\
          -2\alpha\,v_{i-1}+(4\alpha+1)\,v_i-2\alpha\,v_{i+1} & = & 1\\
         \end{array}\right.\label{eqn_appendix_2:3}
\end{equation}

Notice from  Eq.~(\ref{eqn_appendix_2:2}) that $\alpha=0$ means no friction at 
all, and thus, the individuals are allowed to move free from drag. It can be 
verified that $v_i=1$ solves the set (\ref{eqn_appendix_2:3}) for this 
scenario. The $\alpha=0$ scenario is expected to occur, however, for densities 
below a contacting threshold. \\

\begin{figure}[htbp!]
\centering
\includegraphics[width=0.7\columnwidth]
{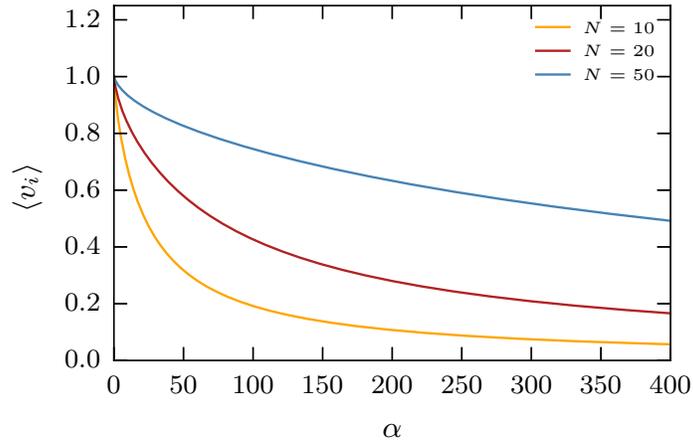}
\caption{\label{fig:appendix_2:1} Mean velocity of the bottom half of the 
individuals vs. the parameter $\alpha$. Both axis are dimensionless. $N$ 
corresponds to the number of individuals. }
\end{figure}

A boundary condition needs to be imposed in order to solve 
Eq.~(\ref{eqn_appendix_2:3}) for $\alpha\neq 0$. We fix 
$v_i=v_{i+1}$ in the middle of the corridor since the velocity profile 
should be symmetrically distributed with respect to the mid-axis of the 
corridor. Fig.~\ref{fig:appendix_2:1} shows the computed mean velocity for the 
bottom side profile as a function of $\alpha$.\\

Fig.~\ref{fig:appendix_2:1} exhibits a decreasing behavior for increasing 
values of $\alpha$. As explained above, the maximum value occurs at $\alpha=0$ 
(\textit{i.e.} $\langle v_i\rangle=1$). However, the decreasing slope slows 
down for an increasing number of individuals. This corresponds to a flattening in 
the velocity profile (see Section \ref{results} for details).  \\

\begin{figure}[htbp!]
\centering
\includegraphics[width=0.7\columnwidth]
{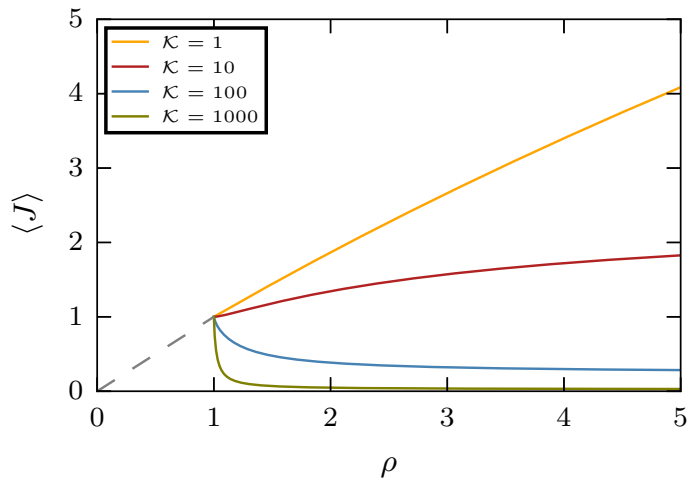}
\caption{\label{fig:appendix_2:2} Mean flux of the bottom half of the 
individuals vs. the pedestrian (global) density $\rho$ (see text for details). 
Both axis are dimensionless. The number of individuals across the 
corridor was set to $N=10$ and the contacting threshold was set to $\rho_0=1$. 
The ``packing coefficient'' was set to $c=1/\mathcal{K}$ (and thus, making the 
term $c\alpha$ independent of friction). The dashed line corresponds to the 
flux at the low density regime (say, $\langle v_i\rangle=1$).   }
\end{figure}

The mean flux of individuals can be built from the mean velocity and the 
corresponding pedestrian density as follows

\begin{equation}
 J=\left\{\begin{array}{lcl}
          \rho & \mathrm{for} & \alpha=0 \\
          & & \\
          (\rho_0+c\,\alpha)\,\langle v_i\rangle & \mathrm{for}  & 
\alpha>0\\
         \end{array}\right.\label{eqn_appendix_2:4}
\end{equation}

where $\langle v_i\rangle$ equals unity for the case $\alpha=0$, and thus, it 
was omitted in (\ref{eqn_appendix_2:4}). The density $\rho=\rho_0+c\,\alpha$ 
corresponds to the packing density (that is, the density above the contacting 
threshold) and $c$ corresponds to a somewhat ``packing coefficient''. 
Fig.~\ref{fig:appendix_2:2} shows the flow as a function of the density, 
assuming $\rho_0=1$ for simplicity. \\

The pedestrian flux $J$ attains two possible behaviors, according to 
Fig.~\ref{fig:appendix_2:2}. For packing coefficients $c<0.05$, the flux 
diminishes as the corridor becomes more crowded. But, if $c$ surpasses this 
threshold, the flux slope becomes positive, although the mean velocity 
diminishes. We conclude that the role of the pedestrians' friction coefficient is crucial 
for building the fundamental diagram.

\subsection{Friction modification}

As already mentioned, the results shown so far indicate that friction may be the 
key magnitude for fitting the fundamental diagram into the experimental data. We 
want to make clear that fitting the experimental data means mimicking 
(qualitatively) the congested regime reported by different authors (including 
the Jamaraat study in Ref.~\cite{helbing3}) for corridors as width as 22~m. The 
original version of the SFM proposes the same 
friction coefficient for the pedestrian-pedestrian interaction and the 
pedestrian-wall interaction. The proposed value was $\kappa = 2.4\times10^{5}$. 
This value is widely used in many studies.  \\

\begin{figure}[htbp!]
\centering
\includegraphics[width=0.7\columnwidth]
{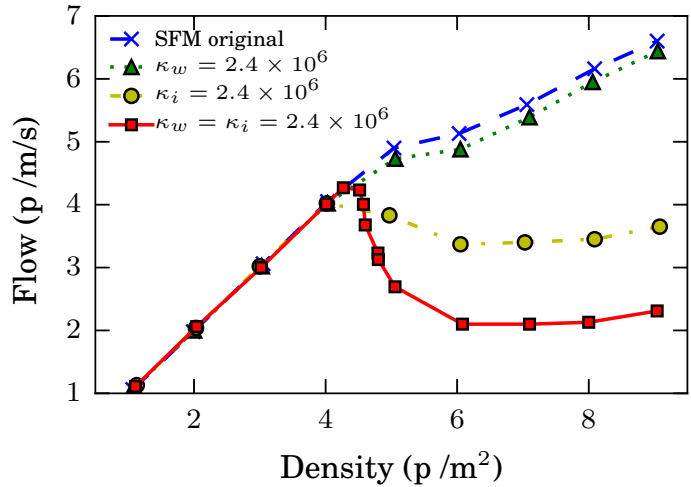}
\caption{\label{fgmodified-w22} Fundamental diagram (flow vs. density) for different friction coefficient (see legend for the corresponding values). The simulated corridor was 28~m length. Pedestrians walk from left to right with periodic boundary condition in the $x$-direction. Initially, pedestrians were randomly distributed. For each density, we measure the flow once the system reaches the stationary state.}
\end{figure}

We tested the friction coefficient modification in Section \ref{appendix_2} and we found that the fundamental diagram experiences a qualitatively change when the friction coefficient $\kappa$ is varied. We further performed numerical simulations in the context of the SFM. We call $\kappa_i$ as the friction coefficient of the pedestrian-pedestrian interaction and $\kappa_w$ as the friction coefficient of the pedestrian-wall interaction. Fig.~\ref{fgmodified-w22} shows the flow vs. density for different values of $\kappa_i$ and $\kappa_w$.\\

The triangular symbols in Fig.~\ref{fgmodified-w22} correspond to the increase in one order of magnitude of the wall friction (now $\kappa_w = 2.4\times10^{6}$), leaving the pedestrian-pedestrian friction unchanged (\textit{i.e.}, $\kappa_i = 2.4\times10^{5}$). We can see that the flow reduces a little bit, but this is not enough to change significantly the congested regime. \\

The circles in Fig.~\ref{fgmodified-w22} correspond to a modification of the friction between pedestrians without changing the value of the wall friction. We increased the pedestrian-pedestrian friction by a factor of ten ($\kappa_i = 2.4\times10^{6}$). Here we see a significant reduction of the flow. The qualitative behavior resembles the fundamental diagram reported by Helbing \textit{et al}~\cite{helbing3}. with a well defined congested regime for the greatest densities.\\

We also tested the case were both friction coefficients surpass ten times the value of the original model (now $\kappa_w = \kappa_i = 2.4\times10^{6}$). The squared symbols represent this scenario. As expected, the flow reduces significantly with respect to the original case (cross symbol). Interestingly, the reduction of the flow is more than the reduction due to the increment of $\kappa_i$ plus the reduction of the flow due to $\kappa_w$. This behavior indicates that the superposition principle does not hold in this system because of the non-linearity of the equation of motion.     \\

This finding allows us to affirm that the friction plays a crucial role in the functional behavior of the fundamental diagram. The increment of both individual-individual friction and wall friction are determinant in order to achieve a congested regime. More specifically, the empirical behavior for the fundamental diagram can be achieved by properly increasing the friction coefficients. In \ref{appendix_3} we show that the friction modification does not alter already studied behaviors of pedestrian dynamics.\\

Recall that other authors address the ``congested regime problem" by modifying different aspects of the model. Ref.~\cite{parisi2} imposes zero desired velocity once pedestrians are close enough, Ref.~\cite{johansson} increases the relaxation time in order to slow down the net-time headway, and more recently,  Ref.~\cite{kwak} induce the jamming transition by an attraction. Many of these approaches seem to be equivalent. In Section \ref{appendix_1} we discuss how the modification of the relaxation time and the increment of the friction coefficient yield a similar effect since both affect the same term in the reduced-in-units equation of motion.  The relaxation time, however, also affects the social interactions (see Sections \ref{Hypotheses} and \ref{appendix_1}).   \\

We claim that in real scenarios, a combination of all these factors may be the cause of the marked flow reduction that portrays the fundamental diagram. The pedestrians path can be very complex even if it is a simple enclosure (straight corridor) and the target is well defined (unidirectional flow). Beyond the complexities given by the internal motivations of pedestrians, we strongly suggest studying and modeling coefficients of friction between individuals and the friction with the walls. These two parameters have shown to be very important in the pedestrian dynamics and deserve a closer inspection in future research.\\

We want to emphasize that the proposals stated in 
Refs.~\cite{kwak,parisi2,johansson}  only apply under normal conditions. 
If a crowd is under high levels of anxiety, pedestrians will neither keep 
distance between each other nor will feel the urge to see an ``attraction". 
Thus, studying the friction coefficients may be a critical factor to properly 
reproduce the dynamics of a massive evacuation under stress.\\

With all these insights, we can say that the narrow corridors have no drawbacks in the fitting of the flow vs. density because very high velocities are not attainable. This happens because, in narrow corridors, the friction of the walls has a lot of ``relative weight" in the overall friction of the system. 
The friction exerted by the walls is fundamental in order to produce the parabolic shape of the velocity profile. The walls provide friction force in the opposite direction to the speed of the individual (drag backwards), since they act like a fixed pedestrian. In other words, friction between pedestrians can produce either drag forward or drag backwards depending on the contacting pedestrians velocities (see Eq.~\ref{granular}).\\

In this subsection, we have shown that an adequate modification of the friction coefficients yields a fundamental diagram that follows qualitatively the behavior reported through empirical data (say flow reduction for the highest densities). We have also discussed different approaches proposed by other authors in order to overcome this problem. See Section \ref{appendix_1} for a more detailed discussion.\\

\subsection{\label{clusters}Clusters}

\begin{figure}[!htbp]
\centering
    \subfloat[]{\includegraphics[width=0.40\columnwidth]{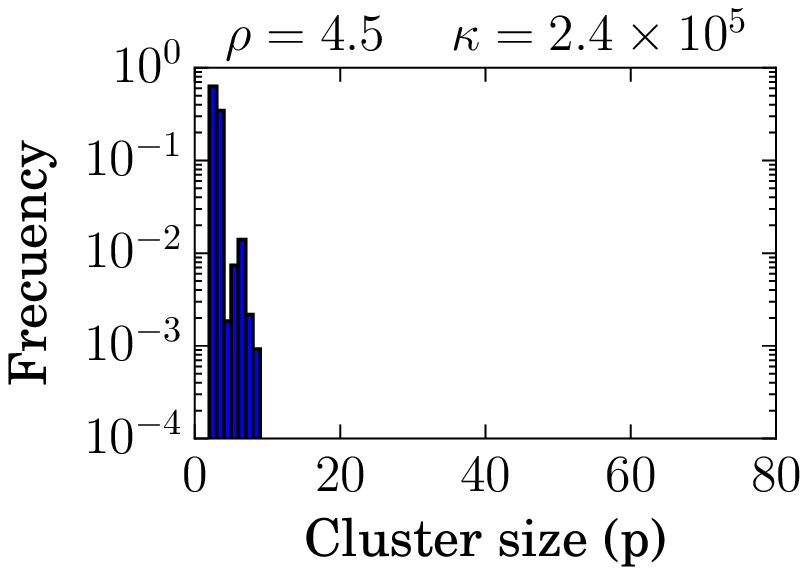}\label{size_distribution_w22_density4.5}}\ 
    \subfloat[]{\includegraphics[width=0.40\columnwidth]{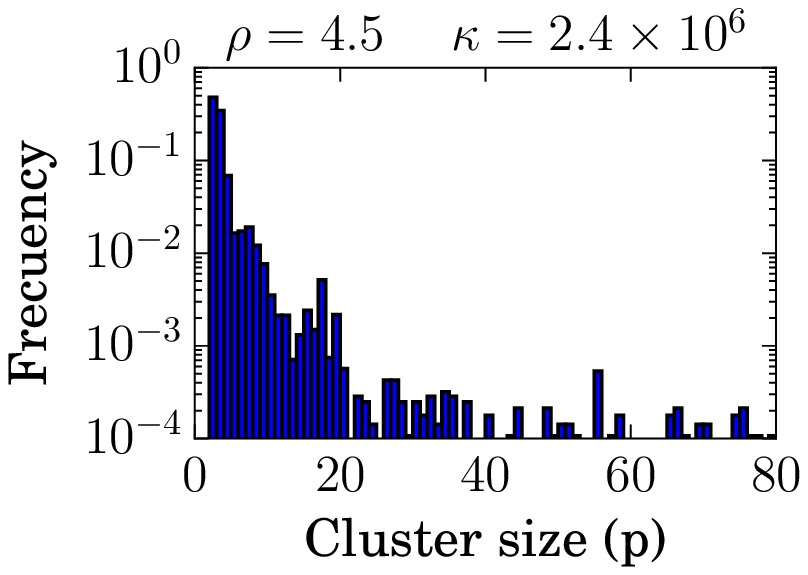}\label{size_distribution_w22_density4.5_kx10}}\\
    \subfloat[]{\includegraphics[width=0.40\columnwidth]{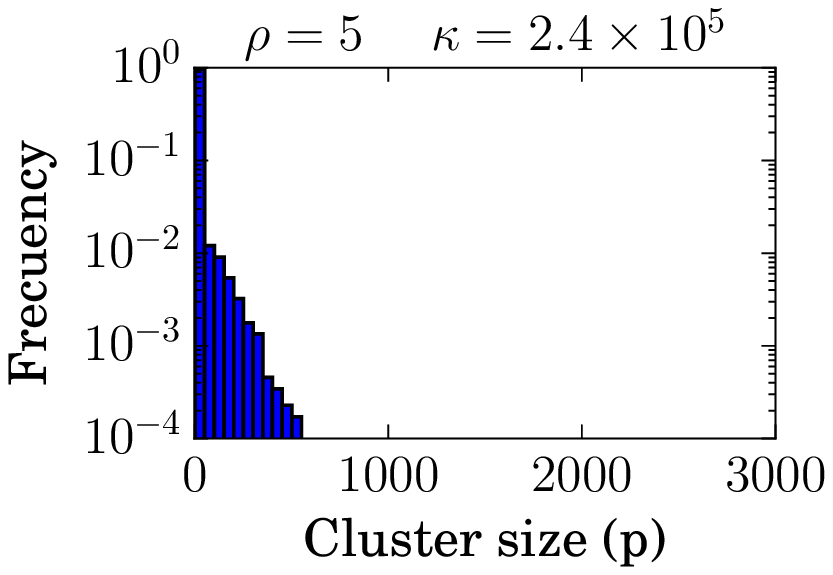}\label{size_distribution_w22_density5}}\
    \subfloat[]{\includegraphics[width=0.40\columnwidth]{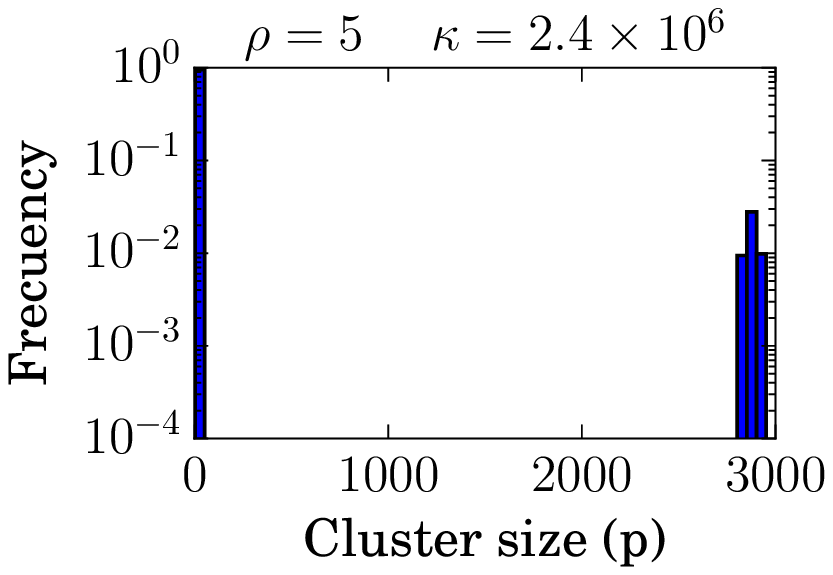}\label{size_distribution_w22_density5_kx10}}\\
    \subfloat[]{\includegraphics[width=0.40\columnwidth]{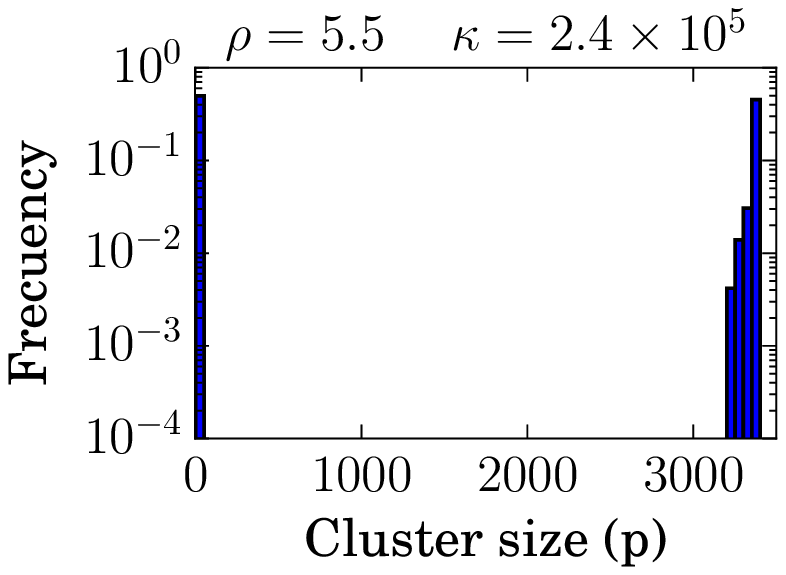}\label{size_distribution_w22_density5_5}}\
    \subfloat[]{\includegraphics[width=0.40\columnwidth]{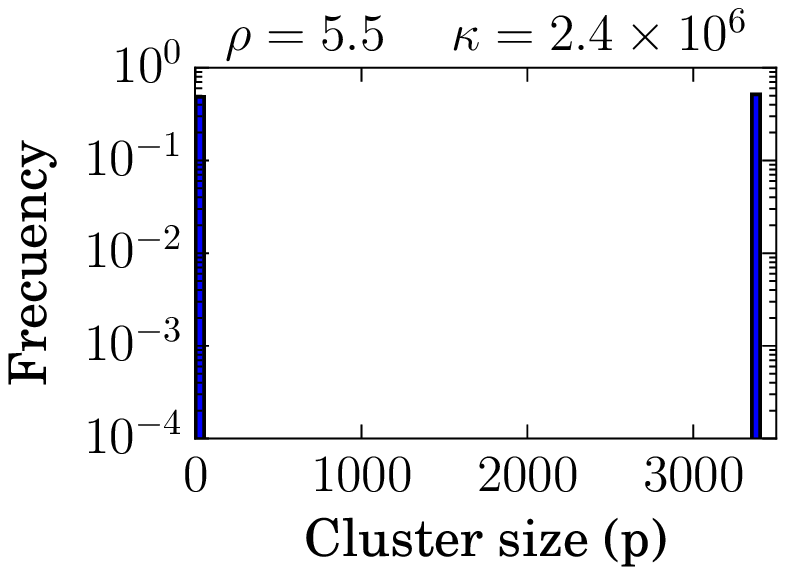}\label{size_distribution_w22_density5_5_kx10}}
\caption[width=0.47\columnwidth]{Cluster size distribution for six different scenarios. (a) SFM friction parameters and $\rho=4.5$, the bin size is 1 p (b) friction parameters increased by a factor of ten and $\rho=4.5$, the bin size is 1 p (c) SFM friction parameters and $\rho=5$, the bin size is 50 p (d) friction parameters increased by a factor of ten and $\rho=5$, the bin size is 50 p (e) SFM friction parameters and $\rho=5.5$, the bin size is 50 p (f) friction parameters increased by a factor of ten and $\rho=5.5$, the bin size is 50 p.}
\label{cluster_distribution}
\end{figure}

Cluster formation is a very important process in pedestrian dynamics. Moreover, it is the key process that explains the clogging phenomena in bottleneck evacuations. We analyzed the clustering formation according to the granular cluster definition given in Section~\ref{granular-cluster}. Fig.~\ref{cluster_distribution} shows the histograms of the cluster size distribution for three different densities, from top to bottom: $\rho=4.5$, $\rho=5$ and $\rho=5.5$. These three
densities are representative of the crossover between
a non-clusterized regime and a unique giant cluster
regime. We studied two situations for each density: the
original SFM on the left hand side plots, and the enhanced
friction situation on the right hand side plots (see the
caption for details).\\

We found two unexpected results. Increasing the density produces bigger size clusters until the size distribution suddenly switches to a bimodal distribution (compare Fig. \ref{size_distribution_w22_density5} with Fig. \ref{size_distribution_w22_density5_5} and Fig. \ref{size_distribution_w22_density4.5_kx10} with Fig. \ref{size_distribution_w22_density5_kx10}). Once
this phenomenon occurs, any of two possibilities may
appear: the pedestrian belongs to a small cluster (out
of many ``caged" in the crowd) or he (she) belongs to the
giant cluster (with a size comparable to the entire crowd).
In other words, the bimodal distribution occurs because,
after the giant component is formed, many small clusters remain ``caged" 
inside the giant component. These small clusters (or single individuals)
do not touch permanently any pedestrian belonging to the giant component. \\

We also observed that this phenomenon is controlled by the friction. For higher frictions, the crossover to the bimodal distribution occurs at lower densities (see Fig.~\ref{size_distribution_w22_density5} and Fig.~\ref{size_distribution_w22_density5_kx10}). Despite the fact that both correspond to the same density, Fig.~\ref{size_distribution_w22_density5_kx10} already attains the bimodal distribution since the friction force is ten times greater than in Fig.~\ref{size_distribution_w22_density5}. This peculiar phenomenon
occurs because in an enhanced friction scenario the
individuals find it harder to detach from each other. On
the opposite, when the friction is weak, the individuals
detach themselves more easily, leading to a situation
where large clusters are less probable.\\

\begin{figure}[htbp!]
\centering
\includegraphics[width=0.7\columnwidth]
{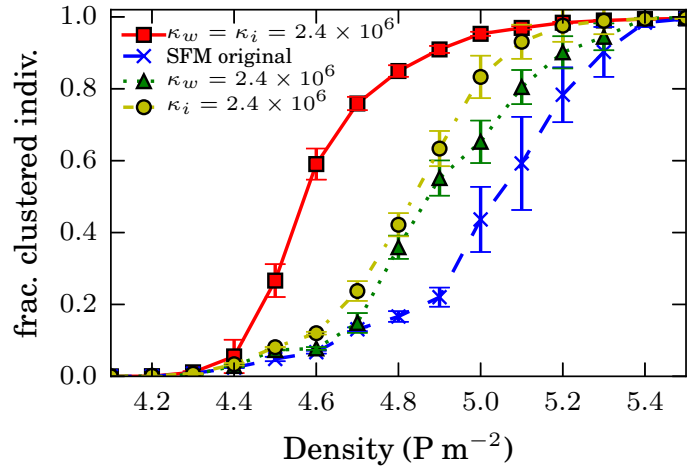}
\caption{\label{fic} Fraction of clustered individuals as a function of the density. Squared symbols correspond to $\kappa_w=\kappa_i=2.4\times 10^6$, circles to $\kappa_i=2.4\times 10^6$ and $\kappa_w=2.4\times 10^5$, triangles to $\kappa_i=2.4\times 10^5$ and $\kappa_w=2.4\times 10^6$ and crosses to the coefficients of the original SFM ($\kappa_i=2.4\times 10^5$ and $\kappa_w=2.4\times 10^5$). The pedestrians walk across a corridor with $v_d=1$m~s$^{-1}$. The measurements were recorded every 0.5~s once the system reached the stationary state. The values corresponding to the fraction of clustered individuals were averaged over 170~s. The cluster cutoff distance was 0.46~m (equivalent to the shoulder width of the pedestrians).}
\end{figure}

To get a better view of this phenomenon, we represent in Fig.~\ref{fic} the fraction of clustered individual as a function of the density for four different situations in which we only changed the friction coefficients. The fraction of clustered individuals is defined as the amount of pedestrians that belong to a cluster (of two or more individuals) over the total number pedestrians in the corridor. 
We can see that there is a transition from a non-clustered crowd to a full-clustered crowd. This transition occurs in a very narrow range of densities, causing the cluster formation process to occur between $4.4<\rho<5.3$. When the friction coefficient is enhanced, the transition takes place at a lower density threshold. This result is in complete agreement with the histograms shown in Fig.~\ref{cluster_distribution}, confirming the fact that friction promotes the formation of clusters.\\

As expected, the transition sharpens when both $\kappa_i$ and $\kappa_w$ are increased (squared symbol in Fig.~\ref{fic}). However, the fraction of clustered individuals is always greater than in the original SFM when just one of the
two coefficients is increased. Notice that there is not a big difference between the modification of $\kappa_i$ and the modification of $\kappa_w$. This suggests that it does not matter if the clusterization starts at the areas close to the walls or in the middle of the crowd.\\

Two main conclusions can be outlined from the above
results. The friction coefficients between the pedestrians
(and with the walls) appear as decisive parameters
with respect to the crossover between the freely moving
regime and the slow down regime. The precise (density) threshold between both regimes will actually depend on
the friction coefficients since these control the efforts
required by the pedestrians to detach from each other.
Above this threshold (say, at high densities) the whole
crowd slows down since the pedestrians appear mostly
clustered or ``caged" in a clustered environment.\\

\section{\label{conclusions}Conclusions}

Our investigation focused on the fundamental diagram in the context of the 
SFM. We compared empirical data recorded at 
the entrance of the Jamaraat bridge (see Ref.~\cite{helbing3}) with our own SFM 
simulations. We observed  that the  SFM, in its 
original version, does not properly reproduce the empirical fundamental 
diagram since the pedestrian flow increases even for highly dense 
crowds. The reasons for this mismatching were studied through 
numerical computations and by a simple theoretical example. We arrived at the 
conclusion that either increasing the friction coefficient or increasing the 
relaxation time it is possible to achieve a non-increasing 
flow in the congested regime of the fundamental diagram. The 
latter has already been explored in Ref.~\cite{johansson} and a 
similar idea was introduced in Ref.~\cite{parisi2}. We noticed, 
though, that both approaches are equivalent since both affect 
the reduced-in-units equation of motion in a similar fashion.\\

In order to further explore the effect of the friction term on the 
dynamics of the crowd, we performed numerical simulations increasing the 
value of $\kappa$. We were able to reproduce the empirical 
fundamental diagram for sufficiently high values of $\kappa$, 
while keeping the original SFM unchanged (without incorporating neither additional forces nor extra parameters). 
This is actually the main achievement of our investigation. \\

We further explored the velocity profile 
across a corridor. It appears to follow a parabolic-like 
function. The pedestrians at the middle of the 
corridor attain the maximum velocity, while those close to the 
walls attain the minimum. We noticed that the 
velocity profiles, after been scaled by the maximum velocity $v_{max}$ and the 
corridor width $w$, yield a somewhat universal behavior, 
regardless of the corridor width. Thus, we worked on the 
hypothesis that the dynamics should be essentially the same for narrow or wide 
passageways. \\

The presence of clustering structures was found to be controlled 
by the friction coefficient. Interestingly, increasing the 
pedestrian-pedestrian friction ($\kappa_i$) or the pedestrian-wall friction 
coefficient ($\kappa_w$) yields a similar clusterization dependence with 
density.\\

All these phenomena suggest that further research needs to be  
done regarding the friction coefficient. The explored values 
introduced through the investigation, however, should not be considered as 
``empirical'' ones. Its true meaning (within the context of the SFM) is related 
to the other parameters in the model (see Section \ref{appendix_1}). A real consensus 
on empirical values of $\kappa$ is still missing, to our knowledge. Further analyses are needed to fully explore this issue and are currently under development. \\

We further proposed modeling the pedestrian-wall friction 
interaction with a different coefficient than the pedestrian-pedestrian friction 
interaction. This does neither mean a ``re-calibration'' (for 
highly dense crowds) nor a departure from the original SFM model. 
We actually find no reason for changing other parameters, regardless of any 
specific situation. We also stress the fact that studying the friction 
coefficients may be a critical factor to properly reproduce the dynamics of a 
massive evacuation under high levels of anxiety. 
\\  

\section*{Acknowledgments}
This work was supported by the National Scientific and Technical 
Research Council (spanish: Consejo Nacional de Investigaciones Cient\'\i ficas 
y T\'ecnicas - CONICET, Argentina) grant Programaci\'on Cient\'\i fica 2018 (UBACYT) Number 20020170100628BA.

\appendix


\section{\label{appendix_3} Testing out previous results}

In this Appendix, we show that the friction modification does not alter already 
studied behaviors of pedestrian dynamics (see 
Refs.~\cite{Dorso2,Dorso1,Dorso3}). In order to check this out, we performed 
numerical simulations of a room of 225 pedestrians escaping through a narrow door 
(bottleneck enclosure). The evacuation time (clearance time) $t_e$ is a function 
of the desired velocity $v_d$. This function has a minimum such that below it, 
$t_e$ is a decreasing function of $v_d$ while above, the tendency is reversed. 
It means that the evacuation process is optimum at moderated $v_d$. This is a 
well known phenomenon called the Faster-is-Slower effect. This effect was 
reported in numerical simulations \cite{Dorso1,Dorso5} and it has also 
experimental support \cite{parisi1}. \\

\begin{figure}[htbp!]
\centering
\includegraphics[width=0.7\columnwidth]
{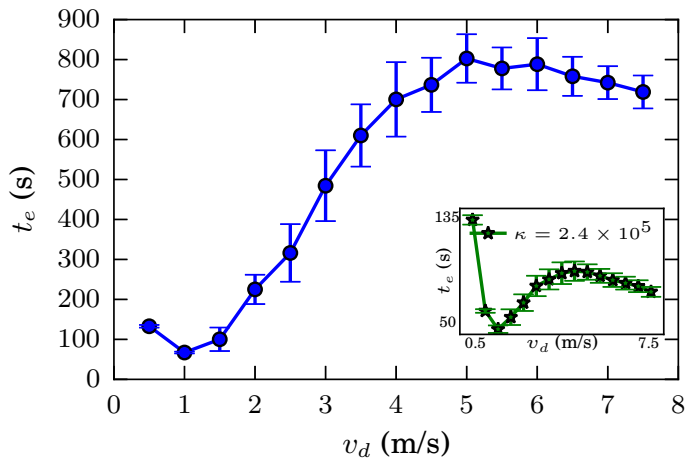}
\caption{\label{fis} Mean evacuation time (seconds) vs. the pedestrian’s desired velocity (m~s$^{-1}$). The room was 20 m $\times$ 20 m size. Mean values were computed from 30 evacuation processes. The pedestrians were initially placed in a regular square arrangement along the room with random velocities, resembling a Gaussian distribution with null mean value. The friction coefficient was $\kappa=2.4\times10^{6}$. The inset shows the evacuation time vs. the desired velocity for the friction coefficient corresponding to the original model ($\kappa=2.4\times10^{5}$).  }
\end{figure}

In Fig.~\ref{fis} we show the evacuation time as a function of the desired velocity corresponding to a friction coefficient ten times greater than the friction coefficient of the original SFM (\textit{i.e.} now $\kappa=2.4\times 10^{6}$). We can see that this modification preserves the Faster-is-Slower effect (and the Faster-is-Faster effect). The inset exhibits the evacuation time as a function of the desired velocity with the friction coefficient of the original model ($\kappa=2.4\times 10^{5}$).\\

Another pedestrian dynamics phenomenon is the lane formation reported in 
bidirectional flows \cite{helbing5,feliciani1,guo1,qiao1}. This means, 
individuals spontaneously organize in lanes of uniform walking direction if the 
pedestrian density is high enough. One of the major achievements of the 
SFM is its ability to reproduce this emergent phenomenon in which the 
microscopic interactions between pedestrians is enough to produce a global 
pattern. We run a bidirectional flow simulation (not shown) and verified that 
the friction increment still reproduces the lane formation process.\\  

\begin{figure}[htbp!]
\centering
\includegraphics[width=0.7\columnwidth]
{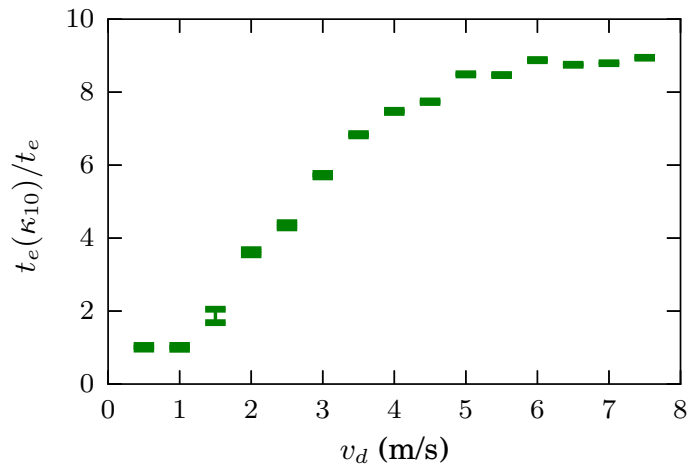}
\caption{\label{rate_fis} Rate of evacuation time ($t_e(\kappa_{10})/t_e$) vs. the desired velocity. The room was 20 m $\times$ 20 m size. Mean values were computed from 30 evacuation processes. The pedestrians were initially placed in a regular square arrangement along the room with random velocities, resembling a Gaussian distribution with null mean value. The friction coefficient were $\kappa=2.4\times10^{6}$ and $\kappa=2.4\times10^{5}$. }
\end{figure}

Fig.~\ref{rate_fis} exhibits the ratio between the evacuation time for the 
friction modified model and the evacuation time for the original model. This 
ratio is shown as a function of the desired velocity. As the desired velocity 
increases, the rate of evacuation times approaches a constant value $\sim$ 9. 
This result is in agreement with the formula Eq.~(9) in Ref.~\cite{sticco}. The 
friction force dominates in the range of high desired velocities since it 
promotes the collisions between pedestrians. Within this regime, the Eq. (9) of  
Ref.~\cite{sticco} becomes a good approximation of the evacuation time. Thus, it 
is reasonable that dividing $t_e(\kappa_{10})$ (for the enhanced friction) by $t_e$ (for the 
original model) yields a constant value for high $v_d$.  \\ 

We also tested the situation in which $k_i=2.4\times 10^{6}$ and $k_w=0$. We obtained similar results to the ones obtained when $k_i=k_w=2.4\times 10^{6}$ because the friction of the walls in bottleneck evacuations do not play a fundamental role in the evacuation process. 










\end{document}